\documentclass[graybox]{svmult}

\usepackage{graphicx}
\usepackage{bm}
\usepackage{amsmath}
\usepackage{amssymb}
\usepackage[colorlinks=true, linkcolor=blue, citecolor=red, urlcolor=blue]{hyperref}
\usepackage{multicol}           
\usepackage[bottom]{footmisc}   

\makeindex             

\begin{document}

\title*{Scaling properties in the adsorption of ionic polymeric surfactants on generic nanoparticles of metallic oxides by mesoscopic simulation}

\titlerunning{Scaling properties for ionic surfactants}

\author{E. Mayoral and E. Nahmad-Achar}

\institute{Estela Mayoral \at Instituto Nacional de Investigaciones Nucleares,\\
Carretera M\'exico-Toluca s/n, 52750 Edo.\!\! de Mex., M\'exico\\
\email{estela.mayoral@inin.gob.mx}
\and Eduardo Nahmad-Achar \at Instituto de Ciencias Nucleares, Universidad Nacional Aut\'onoma de M\'exico,\\
Apartado Postal 70-543, 04510 M\'exico DF, M\'exico\\
\email{nahmad@nucleares.unam.mx}}

\thispagestyle{empty}
\maketitle

\setcounter{page}{1}


\abstract{We study the scaling of adsorption isotherms of polyacrylic dispersants on generic surfaces of metallic oxides $XnOm$ as a function of the number of monomeric units, using Electrostatic Dissipative Particle Dynamics simulations. The simulations show how the scaling properties in these systems emerge and how the isotherms rescale to a universal curve, reproducing reported experimental results. The critical exponent for these systems is also obtained, in perfect agreement with the scaling theory of deGennes. Some important applications are mentioned.}

\section{Introduction}

Polyelectrolyte solutions have properties quite different from those observed in solutions of uncharged polymers, and their behavior is less well known~\cite{odjik,dobrynin,deGennes}. In particular, the scaling of some quantities could present a different behavior and atypical scaling exponents could be found. In most cases, the statistical properties of these interesting systems cannot be obtained analytically because of the long-range Coulombic repulsion produced by the presence of small mobile counterions in the bulk, which interact both with the charge in the polymer and with each other. The use of simulation methodologies have shown, however, to be a promising tool in the study of very complex systems~\cite{fermeglia}.

In our case of study, the presence of big charged molecules (such as polymers) and small ones (like counterions and solvents) involving different length and time scales, makes an electrostatic mesoscopic approach a good alternative.  One of these mesoscopic approaches is {\it Dissipative Particle Dynamics} (DPD), which is a Langevin dynamics approximation where in the simulation the fluid is represented by virtual particles which interact with each other through three forces: conservative, random and dissipative. The conservative force includes repulsive and electrostatic interactions, and determines the equilibrium state of the system, whereas the dissipative and random forces act as a thermostat and they allow transport properties preserving the thermodynamic equilibrium. The electrostatic interactions in DPD simulations were first incorporated by Groot~\cite{groot2}, who solved the electrostatic field locally on a lattice. An alternative way to solve the electrostatic problem in DPD was later developed by Gonzalez-Melchor et al.~\cite{melchor}, where the calculation of the electrostatic interactions employs the standard Ewald sum method and, in order to prevent the artificial ionic pair formation, charge distributions on DPD particles were included.

In this work we study, through electrostatic mesoscopic dissipative particle dynamics simulations, the adsorption of dispersants onto pigments and show the obtained density profiles, the adsorption isotherms, and their scaling properties.

\section {Mesoscopic Approach}

One of the main problems in many areas of industrial and academic interest is that the systems that one deals with are constituted by many particles of different length scales, interacting in different time scales. In order to simplify the study of these systems, in the early 1990s Hoogerbruge and Koelman~\cite{hoogerbruge} introduced a mesoscopic simulation technique. This is known as Dissipative Particle Dynamics (DPD) and is a coarse graining approach which consists of representing complex molecules as soft spherical beads interacting through a simple pair-wise potential, and thermally equilibrated through hydrodynamics~\cite{groot}. In this formalism, the beads follow Newton´s equations of motion

\begin{equation}
	\frac{dr_i}{dt} = v_i, \qquad \frac{dv_i}{dt} = f_i
\end{equation}

\noindent where $_ri$ and $v_i$ are the position and the velocity of the $i$-th particle respectively, and the force $f_i$ is constituted by three components

\begin{equation}
f_i=\sum_j\, \left( f^C{}_{ij} + f^D{}_{ij} + f^R{}_{ij} \right)
\end{equation}

\noindent corresponding to conservative, dissipative and random contributions respectively. The sum runs over all neighboring particles within a certain distance $R_c$. The conservative force $f^C$ derives from a soft interaction potential and there is no hard-core divergence as in the case of a Lennard-Jones potential, thus providing a more efficient scheme of integration; it has the form

\begin{equation}
f^C{}_{ij} = a_{ij}\, \omega^C(r_{ij})\, \frac{r_{ij}}{|r_{ij}|}
\end{equation}

When we need to introduce a more complex molecule such as a polymer we use beads joined by springs, so we also have an extra spring force given by $f^S{}_{ij} = k\, r_{ij}$ if $i$ is connected to $j$. The dissipative $f^D$ and random $f^R$ standard DPD forces are given by
\begin{equation}
f^D_{ij} = -\gamma\,\omega^D(r_{ij})\, \frac{(r_{ij} \cdot v_{ij})\, r_{ij}}{|rij|^2}
\end{equation}

\noindent and

\begin{equation}
f^R{}_{ij} = -\sigma\,\omega^R(r_{ij})\, \frac{\theta_{ij}}{\delta_t^{1/2}}\, \frac{r_{ij}}{|r_{ij}|}
\end{equation}

\noindent Here, $\delta_t$ is the time step; $v_{ij}=v_i-v_j$ is the relative particle velocity; $\theta{ij}$ is a random Gaussian number with zero mean and unit variance; $\gamma$ and $\sigma$ are the dissipation and the noise strengths respectively; and $\omega^C(r_{ij}),\ \omega^D(r_{ij})$ and $\omega^R(r_{ij})$ are dimensionless weight functions. Not all these quantities are independent: some of them are related through the fluctuation-dissipation theorem~\cite{warren} by $\gamma = \sigma^2/2\kappa_B T$ and $\omega^D(r_{ij}) = [\omega^R(r_{ij})]^{1/2}$, with $\kappa_B$ the Boltzmann constant and $T$ the temperature.

The methodology used in our mesoscopic simulations and, specifically, the electrostatic DPD methodology, is briefly described in the following subsection.

\subsection{Mesoscopic simulation: electrostatic dissipative particle dynamics}

\begin{figure} 
\begin{center}
\scalebox{0.6}{\includegraphics{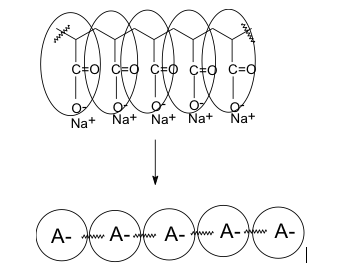}}
\end{center}
\caption{Mesoscopic identification for a polyelectrolyte such as the sodium salt of $PAA$.}
\label{MapeoDPD}
\end{figure}

We consider in our study an ionic polymeric dispersant, for example polyacrylic acid ($PAA$) or a salt derived from it, in water, and in the presence of substrate particles which we will take to be metallic oxides, such as $TiO_2$, $Al_2O_3$, $CeO_2$, etc. We map the polymer chain into beads which we will call {\it DPD beads} as shown by the label $A-$ in Figure~\ref{MapeoDPD}. Each DPD bead has a volume $v_{DPD} = 90$ \AA$^3$ and radius $r_{DPD} = 2.78$ \AA, which corresponds with the volume of three water molecules. We can represent a $PAA$ chain by $N_{DPD}$ beads of carboxylate monomeric units joined by springs with some spring constant $k$; in this case $N_{DPD} = v_{mon} N/ v_{DPD}$ where $v_{mon}$ is the volume of a carboxylate monomeric unit and $N$ the number of monomeric units in the chain.

As $PAA$ and its salt derivatives tend to be very hydrophilic, the adsorbed segments see the substrate as a flat surface when in a good solvent. If $R$ is the effective radius of the substrate particle and $R_g = a_f N^\nu$ is the radius of gyration of the polymer chain, with $a_f{}^3$ proportional to the Flory volume, this means that $R > a_f\,N^{3/5}$. For $CeO_2$ and $Al_2O_3$ nanoparticles, for example, we have $R_{CeO_2} \simeq 10$ nm and $R_{Al_2O_3} \simeq 20$ nm which would give us, for $PAA$ and its salt derivatives, $N < 400$ ($Mw < 40,000$ gr/mol) for $CeO_2$ and $N < 1250$ ($Mw < 125,000$ gr/mol) for $Al_2O_3$. For $TiO_2$ we have even larger radii. This accommodates even the higher molecular weight dispersants, so that a flat substrate approximation is appropriate in our mesoscopic approach.

As mentioned in the Introduction, we here replace the point charge at the center of the DPD particle by a charge distribution throughout the particle. This is in order to avoid the formation of artificial clusters from oppositely charged ions. Groot~\cite{groot2} solved the problem by the calculation of the electrostatic field on a grid. The algorithm is known as the particle-particle-particle mesh (PPPM) algorithm, but in the original version of the PPPM algorithm the far field was solved using Fourier transforms. In~\cite{melchor} one of us and coworkers solved this problem by combining the standard method with charge distributions on particles, adapting the standard Ewald method to DPD particles. In the present work we use the latter method because the Ewald sum technique is the most employed route to calculate electrostatic interactions in microscopic molecular simulations. We take, as in~\cite{melchor}

\begin{equation}
\omega^C(r) = \omega^R(r) = \omega^D(r)^{1/2} = \omega(r)
\end{equation}

\noindent with

\begin{equation}
\omega(r) = \begin{cases} 1 - r/R_c &:\quad r \leq R_c \\
 0 &: \quad r > R_c
\end{cases}
\end{equation}

\noindent being $R_c$ the cutoff distance, taken here to be $6.46$ \AA\ (the simulation characteristic length). We also take $\sigma=3$. We represent the $PAA^{N-}$ with $N$ DPD beads, each one of $v_{mon} = 90$ \AA$^3$, bonded by a spring with $k=100$. The $Na^+$ ions were simulated by one DPD bead each with charge $1^+$, and $3$ water molecules per neutral DPD particle. These values reproduce the isothermic compressibility of water in standard conditions.  All quantities are reduced adimensional quantities. $\kappa_B T = 1$, the adimensional integration step taken is $\Delta t^*=0.02$, and the average total density is $\rho^*=3$.

\section{Results and Discussion}

\subsection{Results for adsorption isotherms}

DPD electrostatic simulations were performed using our mesoscopic model described in the last section, in order to obtain the adsorption isotherms for $[PAA^{N-}][Na^+]_N$ on generic surfaces of metallic oxides $XnOm$ at a basic pH. The length for the $PAA$-DPD molecule was varied as $N = 2, 4, 8, 16$ and $32$ DPD particle units. The repulsive constants $a_{ij}$ in the DPD model were considered as $a_{W-PAA^-}=100$, $a_{W-Na^+}=100$, $a_{W-H_2O}= 100$,  $a_{H_2O-PAA^-}= 82$, $a_{H_2O-Na^+}=25$, $a_{PAA^--Na^+}=25$. These values can be obtained from solubility parameters, and a more refined calculation can be made by using affinity parameters (cf.~\cite{jcp}).

\begin{figure}
\begin{center}
\scalebox{0.53}{\includegraphics{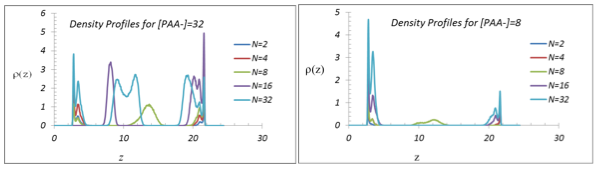}}
\end{center}
\caption{
(Color online) Density profiles for $[PAA]=32,\ \hbox{and}\ 8$.}
\label{DensityProfiles}
\end{figure}

The density profiles $\rho(z)$ obtained, describing the spatial organization of the molecules as a function of one of the spatial coordinates, are presented in Figure~\ref{DensityProfiles}. They show that larger molecules tend to adsorb at the edges of the box (which represent the metallic substrate), and remain less in the aqueous medium (in between the box walls), where smaller molecules can be found.

\begin{figure}
\begin{center}
\scalebox{0.47}{\includegraphics{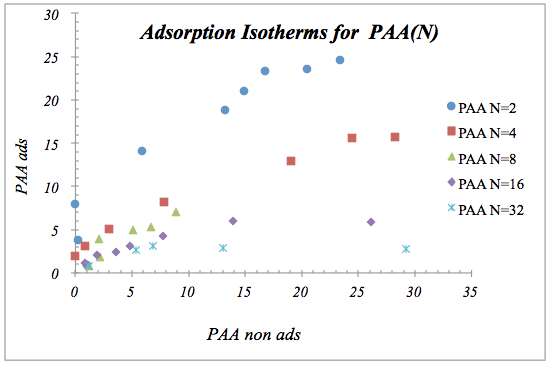}}
\end{center}
\caption{
(Color online) Adsorption isotherms for $PAA$ on a $TiO_2$ surface for different $N$.}
\label{AdsorptionIsotherms}
\end{figure}

To obtain the adsorption isotherms we calculate the amount of polyelectrolyte $\Gamma$ carried by the particle, by the integrating the density profile according to

\begin{equation}
\Gamma = \int^{L_z}_{0}\, [\rho(z) - \rho_{bulk}]\, dz
\end{equation}

\noindent where $L_z$ is the width of the first adsorbed layer and $\rho_{bulk}$ the bulk density. Figure~\ref{AdsorptionIsotherms} shows the number $\Gamma_{mol}$ of $PAA$- molecules adsorbed on a $TiO_2$ surface {\it vs.} the number $\Gamma^b_{mol}$ of non-adsorbed molecules, considering a single adsorbed layer. As expected, the saturation on the surface is reached earlier for large molecules.

\begin{figure}
\begin{center}
\scalebox{0.47}{\includegraphics{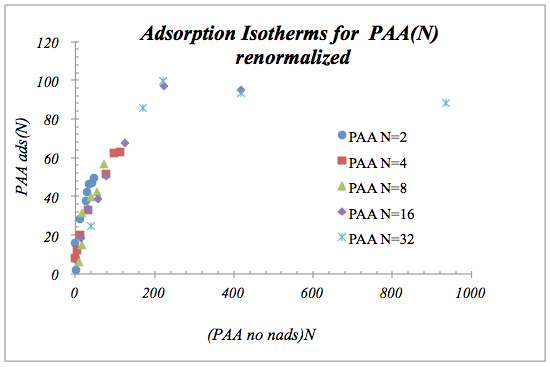}}
\end{center}
\caption{
(Color online) Universal adsorption isotherm for $PAA$ on a $TiO_2$ surface, renormalized.}
\label{AdsorptionIsothermsRenorm}
\end{figure}

\begin{figure}
\begin{center}
\scalebox{0.5}{\includegraphics{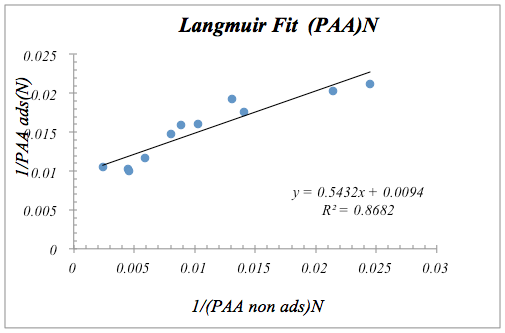}}
\end{center}
\caption{Langmuir fit for the adsorption isotherms of $PAA$ on $TiO_2$.}
\label{LangmuirFit}
\end{figure}

One may easily renormalize these curves, however, by plotting the number of independently adsorbed DPD beads $\Gamma_{DPD}$ {\it vs.} non-adsorbed DPD beads $\Gamma^b_{DPD}$ using $N\,\Gamma_{mol}=\Gamma_{DPD}$. The behavior will then be that of a universal isotherm conformed by the contribution of all sizes, as shown in Figure~\ref{AdsorptionIsothermsRenorm}. Supposing that only one layer is adsorbed on the surface (the self similar region) and that all adsorption positions are equivalent, we can extract the maximum concentration at equilibrium and the adsorption-desorption constant for each isotherm, which is given by the Langmuir isotherm. The dynamic equilibrium is given by $A + N \Leftrightarrow AN$ with velocity constants $K_a$ for the adsorption and $K_d$ for the desorption. The expression for this kind of adsorption model, in the case of {\it neutral} species, is given by the Langmuir isotherm expressed by

\begin{equation}
\frac{1}{\Gamma} = \frac{1}{\Gamma_M} + \frac{1}{\Gamma_M\, K\,C}
\end{equation}

\noindent where $K=K_a/K_d$ and $C$ is the concentration in the bulk ($\Gamma^b$). $\Gamma$ is the adsorbed quantity and $\Gamma_M$ is the maximum adsorbed quantity. The linear fit for this isotherm is shown in Figure~\ref{LangmuirFit}, and is given by $1/\Gamma_M = 0.0094$ and $1/(\Gamma_M\, K) = 0.5432$, from which $\Gamma_M = 106.38\,PAA_{DPD}$ and $K = 0.0173$.

It is interesting that the renormalized behavior adjusts itself to the Langmuir model, {\it even though we are dealing here with charged molecules}.

\subsection{Scaling for $\Gamma_{max}$}

\begin{figure}
\begin{center}
\scalebox{0.5}{\includegraphics{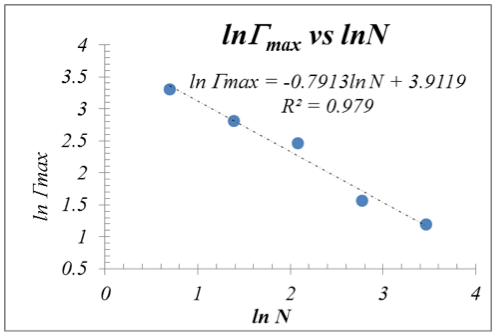}}
\end{center}
\caption{Scaling of $\Gamma_{max}$ with $N$.}
\label{Log-LogGammaMax}
\end{figure}

In the light of the results above, it is interesting to study the behavior of $\Gamma_{max}$ with $N$. This we can do, once more, via DPD electrostatic simulations. $\Gamma_{max}$ is obtained by fitting each isotherm in Figure~\ref{AdsorptionIsotherms} with the Langmuir model, which we have shown to be adequate ({\it vide supra}). Table~\ref{tab1} shows the results for the fit in each case. When we plot $\Gamma_{max}$ {\it vs.} $N$ we obtain the behavior shown in Figure~\ref{Log-LogGammaMax} and the scaling function is $\Gamma_{max} \propto N^{-0.79} \simeq N^{-4/5}$. This result is in perfect agreement with~\cite{deGennes2}.

\begin{table}[h!]
\label{tab1}
\caption{Scaling for $\Gamma_{max}$ as a function of $N$.}
\begin{center}
\begin{tabular}{|c|c|}
	\hline
	$\ln\ N$ & $\ln \Gamma_{max}$\\
	\hline
	\ \ 0.6931\ \  & \ \ 3.3077\ \ \\
	\ \ 1.3863\ \  & \ \ 2.8134\ \ \\
	\ \ 2.0794\ \  & \ \ 2.4581\ \ \\
	\ \ 2.7726\ \  & \ \ 1.5626\ \ \\
	\ \ 3.4657\ \  & \ \ 1.1907\ \ \\
	\hline
\end{tabular}
\end{center}
\end{table}

The scaling theory in the weak adsorption regime indicates that in the flat plateau, i.e., at maximum saturation

\begin{equation}
\gamma_p \sim N^{1/5}
\label{scaling}
\end{equation}

\noindent where $\gamma_p$ is the number of monomers adsorbed in the flat plateau, $\gamma_p = \Gamma_{max}\,N$. Equation (\ref{scaling}) then implies $\Gamma_{max} \sim N^{-4/5} = N^{-0.8}$ which agrees very well with our result.


\section{Discussion}  

Consider a small particle of diameter d1 = 2r1 and a larger one with diameter d2 = 2r2 , i.e., r1<r2 and a1<a2, where a1 and a2 are the corresponding areas. As we saw in the previous section, the number of monomers adsorbed in the flat plateau is γp=ΓmaxN, where Γmax is the number of chains adsorbed, and they scale with N as Γmax~N-4/5 and γp ~N1/5.
Let Γmax be the number of chains of size N per unit area needed in order to cover satisfactorily some amount, say 1 mol, of material (substrate). If we want to cover a surface of area a1, then c =a1Γmax1 chains are needed. Now suppose that the weight of one monomeric unit is 1 unit of mass, then a1Γmax1 [chains] = a1Γmax1N1=a1γp1 and for the same amount of material but with area a2, we will require a2Γmax2 [chains]= a2γp2. Let κ be the amount of mass needed in order to cover the surface of particles of diameter 2r1 divided by the mass of dispersant necesary to cover the surface of particles of diameter 2r2 , then κ= (a1γp1)/ (a2γp2)=(a1N11/5)/(a2N21/5) that is κ = (r2/r1)(N1/N2)1/5.
We can make use of this last expression to analyze different cases:
Case 1. If N1=N2 then κ=r2/r1 and the result obtained in section 2.3 is reproduced. 
Case 2. If we wanted to use the same amount of dispersant, taking dispersants with different length N1 and N2 and having the same chemistry, κ=1 and 1=(r1/r2)(N1/N2)1/5, that is N1=(r1/r2)5N2. We would need a dispersant with a very small degree of polymerization compared with N2 for r1<<r2. In this case the smallest and the best dispersant will be N1=1 (monomeric dispersant) in agreement with the results in ref.[15]. If (r1/r2)5<<1, a change in the chemistry of the dispersant would be a better option.
Case 3. In the limit of a flat approximation we can consider N=(R/af)5/3  and we have κ = (r2/r1)[(r1/r2)5/3]1/5 = (r2/r1)2/3. Taking the values in section 2.3 we obtain κ=(125/20)2/3=3.3993. Comparing this result to our estimation in section 2.3, where 6.25 times the dispersant amount was needed for Al2O3-nanoparticles, we can observe that, if we choose a dispersant with an adequate length N, we would need a much smaller quantity.

\section{Conclusions}

Langmuir isotherms were calculated for polyacrylate dispersants adsorbed on metallic oxides, and their scaling properties as a function of the number of monomeric dispersant units obtained via DPD-simulations. It is interesting that the renormalized behavior of these isotherms adjusts itself to the Langmuir model, even though polyelectrolytes are being considered. The critical exponent was obtained, and this agrees perfectly well with the scaling theory in~\cite{deGennes2}.

The results presented here suggest a methodology for estimating the amount of dispersant necessary in different scenarios, and for better choosing the appropriate dispersants. The particular case of the stabilization of metallic nanoparticles is interesting, as their inclusion in many formulations in order to improve performance properties is presently a major area of research. Problems arise because the dimensions of the nanoparticles and polymeric dispersants are similar, and because of the large total surface area to be covered. However, excessive amounts of any surfactant will cause property degradation of the material, and new specially designed surfactants circumvent the need for large quantities. Here it was shown that our simulation results improve upon the experimental values obtained in~\cite{ECJ}.

\section*{Acknowledgments}
This work was partially supported by DGAPA-UNAM (under project IN102811). Valuable support in computing resources was obtained from DGTIC-UNAM.


\begin{thebibliography}{999}


\bibitem{odjik}
	T. Odijk, Macromolecules {\bf 12}, 688, (1979).

\bibitem{dobrynin}
	A.V. Dobrynin, R.H. Colby and M. Rubinstein, Macromoleules {\bf 28}, 1859 (1995).

\bibitem{deGennes}
	P.G. de Gennes, J. Phys. {\bf 37}, 1445 (1976) .
	
\bibitem{fermeglia}
	M. Fermeglia and S. Pricl, Prog. Org. Coat. {\bf 58}, 187 (2007).

\bibitem{groot2}
	R.D. Groot, J. Chem. Phys. {\bf 118}, 11265 (2003)

\bibitem{melchor}
	M. Gonz\'alez-Melchor, E. Mayoral, M.E. Vel\'azquez and J. Alejandre, J. Chem. Phys. {\bf 125}, 224107 (2006).
	
\bibitem{hoogerbruge}
	P.J. Hoogerbrugge and J.M.V.A. Koelman, Europhys, Lett. {\bf 19}, 155 (1992).
	
\bibitem{groot}
	R.D. Groot and P.B. Warren, Europhys. Lett. {\bf 30}, 191 (1995).

\bibitem{warren}
	P. Espa\~nol and P.B. Warren, Europhys Lett. {\bf 30}, 191 (1995).

\bibitem{jcp}
	E. Mayoral and E. Nahmad-Achar, J. Chem. Phys. {\bf 137} (19), 194701 (2012).

\bibitem{deGennes2}
	P.G. de Gennes, P. Pincus, R.M. Velasco, and F. Brochard, J. Phys. {\bf 37}, 1461 (1976).



\bibitem{AGR}
M. Auboy, J.M. di Meglio and E. Rapha\"el, Europhys. Lett. {\bf 24}, 87 (1993).

\bibitem{AGR2}
M. Aubouy and E. Rapha\"el, Macromolecules {\bf 31}, 4357 (1998).






\bibitem{}
E. Hershkovits, A. Tannenbaum and R. Tannenbaum, J. Phys. Chem. C  {\bf 111}, 12369 (2007).

\bibitem{}
E. Hershkovits, A. Tannenbaum and R. Tannenbaum, J. Phys. Chem. B {\bf112}, 5317 (2008).





\bibitem{}
A. Gama Goicochea, E. Nahmad-Achar and E. P\'erez, Langmuir {\bf 25}, 3529 (2009).

\bibitem{ECJ}
E. Mayoral, E. Nahmad-Achar, and J. Rodriguez; Eur. Coat. J. {\bf 12/2012}, 84 (2012).


\end{thebibliography}
\end{document}